# Exploring Performance, Coherence, and Clocking of Magnetization in Multiferroic Four-State Nanomagnets

Mohammad Salehi-Fashami[1,3*], and Noel D'Souza[2]
[1]Department of Physics and Astronomy, University of Delaware, Newark, DE 19716 USA
[2]Department of Mechanical and Nuclear Engineering, Virginia Commonwealth University, Richmond, VA 23284 USA
[3]Advanced Nanomagnetic Technology LLC, Charlottesville, VA 22903 USA
*Corresponding author: mfashami@udel.edu

Nanomagnetic memory and logic are currently seen as promising candidates to replace current digital computing architectures due to its superior energy-efficiency, non-volatility and propensity for highly dense and low-power applications. In this work, we investigate the use of shape engineering (concave and diamond shape) to introduce biaxial anisotropy in single domain nanomagnets, giving rise to multiple easy and hard axes. Such nanomagnets, with dimensions of ~ 100 nm × 100 nm, double the logic density of conventional two-state nanomagnetic devices by encoding more information (four binary bits: "00","11","10","01") per nanomagnet and can be used in memory and logic devices as well as in higher order information processing applications. We study reliability, magnetization switching coherence, and show, for the first time, the use of voltage-induced strain for the clocking of magnetization in these four-state nanomagnets. Critical parameters such as size, thickness, concavity, and geometry of two types of four-state nanomagnets are also investigated. This analytical study provides important insights into achieving reliable and coherent single domain nanomagnets and low-energy magnetization clocking in four-state nanomagnets, paving the way for potential applications in advanced technologies.

## 1. Introduction

The continued downscaling of conventional transistor-based electronics faces a challenging barrier in the form of increasing energy dissipation. In the quest for alternative paradigms, spin- and nanomagnet-based computing architectures [1]–[5] have emerged as promising candidates. Unlike transistor-based devices, nanomagnets experience a correlated switching of spins [6] and do not suffer from current leakage. As a result, these novel methodologies would not suffer from standby power dissipation and offer substantial benefits such as non-volatility, energy-efficiency, high integration density, CMOS-compatibility, and compact implementation of logic gates.

One of the most important properties of ferromagnetic materials is magnetic anisotropy. This intrinsic property of magnetic materials plays an essential role in magnetoelectric applications such as permanent magnets, information storage media and magnetic recording heads, which require the magnetization to be pinned in a defined direction. In nanomagnets, the magnetic anisotropy also depends on the shape of the nanomagnet and its magnetic properties can be engineered by manipulating the shape of the nanomagnet, with different shapes giving rise to different anisotropic behaviors.
Basic shapes of nanomagnets, such as ellipsoid and rectangular (having uniaxial anisotropy and encoding two states or two binary bits "0" & "1") have attracted a lot of attention for its applications in ultra-low power binary logic [7]–[11] and non-volatile memory applications [12]–[14]. Nanomagnets encoding four states, instead of the conventional two-states, have been theoretically demonstrated to implement Boolean logic [15], [16]. Besides increasing the logic density, this four-state scheme also holds promise for higher order computing applications such as associative memory, neuromorphic computing and image processing [17]. Since nanomagnetic logic devices require accurate propagation of magnetic information along dipole-couple nanomagnets, reliable switching behavior is paramount and has been shown to be dependent on shape geometry, with different shapes playing an important role in the magnetization switching behavior and correlation lengths along an array of nanomagnets [18].

A four-state memory element can be implemented with a magnetostrictive layer (for instance, single-crystal Ni), which would exhibit biaxial magnetocrystalline anisotropy in the (001) plane [41]. Epitaxial films of single-crystal (001) Ni can be grown using molecular beam epitaxy (MBE) [19], [20]. Biaxial anisotropy in magnetic thin-films has also been shown in single-crystal films [21], coupled films [22], double-layer films [23], as well as in a four-pointed star-shaped structure [24], with the latter highlighting the relationship between shape-induced biaxial anisotropy and the geometry of a thin magnetic film element, indicating that in a four-pointed star-shaped structure, the high-energy states occur when the average magnetization, $\vec{M}$, was oriented from tip to tip (along the long dimension), while the low-energy corresponds to $\vec{M}$ pointing diagonally (45°, along the short dimension).

Another technique used to modify a nanomagnet's magnetic anisotropy, similar to shape anisotropy, and termed 'configurational anisotropy', involves creating multiple "easy" axes by introducing small modifications to the uniform magnetization of nanomagnets of a specific symmetric shape [25]–[27]. In experiments conducted by Lambson et al. [28], the effect of configurational anisotropy on the magnetic properties of triangular-, square- and pentagonal-shaped nanomagnets was studied. It was observed that by modifying parameters such as sample thickness and concavity of an indentation introduced along the edges, the direction of the easy axes could be individually adjusted. Consequently, nanomagnetic logic devices requiring energy efficiency and performance reliability could exploit the desirable features of this configurational anisotropy scheme, namely, anisotropy control and the ability to create multiple easy axes. In this study, Terfenol-D (amorphous) is chosen as the magnetostrictive material of our nanomagnets due to its high magnetostriction and magnetomechanical coupling constants, values that are instrumental for the realization of reliable and energy-efficient four-state nanomagnetic devices [10].

This paper is organized as follows: section II discusses the theoretical framework and parameters for studying four-state nanomagnets. Section III examines and presents the various magnetization vector patterns in diamond- and concave-shaped nanomagnets using the micromagnetic simulation code, OOMMF [29]. In this section, we study magnetization coherence in two types of four-state nanomagnets (concave- and diamond-shaped) and investigate the influence of shape, size and thickness as well as magnetization clocking with an applied strain on the four-state nanomagnets. Section IV reviews the results in order to determine the best geometry of the four-state nanomagnets for technological applications and finally, in section V, we present our conclusions.

## 2. Method: Micromagnetic Modeling

In this work, studying shape-engineered four-state nanomagnets, two types of shapes are examined: (i) diamond, and (ii) concave nanomagnets (square nanomagnets with concave grooves in its sides). Nanomagnets with these shapes have been shown to possess a fourfold symmetric anisotropy field [27], [30], [31] due to configurational anisotropy and also demonstrate different micromagnetic switching modes. The schematics of a four-state diamond and concave nanomagnet with their easy and hard axes are illustrated in Fig. 1.

In the following sections, micromagnetic modeling is carried out based on the total Gibbs free energy (Equation 1) of these two nanomagnets. Simulations of the magnetization dynamics are performed using the Object Oriented MicroMagnetic Framework (OOMMF) software [29] in order to explore magnetic moment interactions and magnetization switching in these four-state diamond and concave nanomagnets. Micromagnetics is a continuum theory used to describe the magnetization process within ferromagnetic materials. To study the behavior of these nanomagnets, it is necessary to consider the relevant energy terms such as the exchange energy, magnetostatic anisotropy, stress anisotropy, and external magnetic field.

The total energy of these nanomagnets can be defined for a nanomagnet volume of $\Omega$ as:

$$U_i = \oint \left\{ \begin{array}{l} A\left[(\nabla m_x)^2 + (\nabla m_y)^2 + (\nabla m_z)^2\right] - \frac{1}{2}\mu_0 \vec{H}_d \cdot \vec{M} + \\ \left[ \begin{array}{l} B_1(\alpha_1^2 \varepsilon_{xx} + \alpha_2^2 \varepsilon_{yy} + \alpha_3^2 \varepsilon_{zz}) + \\ B_2(\alpha_1 \alpha_2 \varepsilon_{xy} + \alpha_2 \alpha_3 \varepsilon_{yz} + \alpha_3 \alpha_1 \varepsilon_{xz}) \end{array} \right] - \mu_0 \vec{M} \cdot \vec{H} \end{array} \right\} d\Omega \quad (1)$$

In equation 1, the first term represents the exchange energy ($E_{exchange}$) having an exchange constant, A. The second term, $E_{ms}$, denotes the magnetostatic energy of the nanomagnet while $E_{me}$ is the magnetoelastic energy of the magnetostrictive material having magnetoelastic coupling constants, $B_i$, and direction cosines, $\alpha_i$, while experiencing a strain $\varepsilon_{ij}$. The final term, $E_{Zeeman}$, represents the energy of interaction with an external magnetic field, $H$. In this work, magnetocrystalline anisotropy is neglected as the nanomagnet is assumed to have random polycrystalline orientation.

The detailed analytical expressions for exchange energy and shape anisotropy for fourfold square nanomagnets have been investigated with perturbation theory [32]. The magnetization dynamics of a nanomagnet under the influence of an effective field, $\vec{H}_{eff}$, is described by the Landau-Lifshitz-Gilbert (LLG) equation [33]:

$$\frac{d\vec{M}(t)}{dt} = -\gamma \vec{M}_i(t) \times \vec{H}_{eff}^i(t) - \frac{\alpha \gamma}{M_s}\left[\vec{M}_i(t) \times \left(\vec{M}_i(t) \times \vec{H}_{eff}^i(t)\right)\right]$$

(2)

Here, $\vec{H}_{eff}^i$ is the effective magnetic field on the nanomagnet, defined as the partial derivative of its total potential energy ($U_i$) with respect to its magnetization ($\vec{M}_i$), $\gamma$ is the gyromagnetic ratio, $M_s$ is the saturation magnetization of the magnetostrictive layer and $\alpha$ is the Gilbert damping factor [34] associated with internal dissipation in the magnet owing to the magnetization dynamics. Accordingly,

$$\vec{H}_{eff}^i(t) = -\frac{1}{\mu_o \Omega} \frac{\partial U_i(t)}{\partial \vec{M}_i(t)} = -\frac{1}{\mu_o \Omega M_s} \nabla_{\bar{m}} U_i(t)$$

(3)

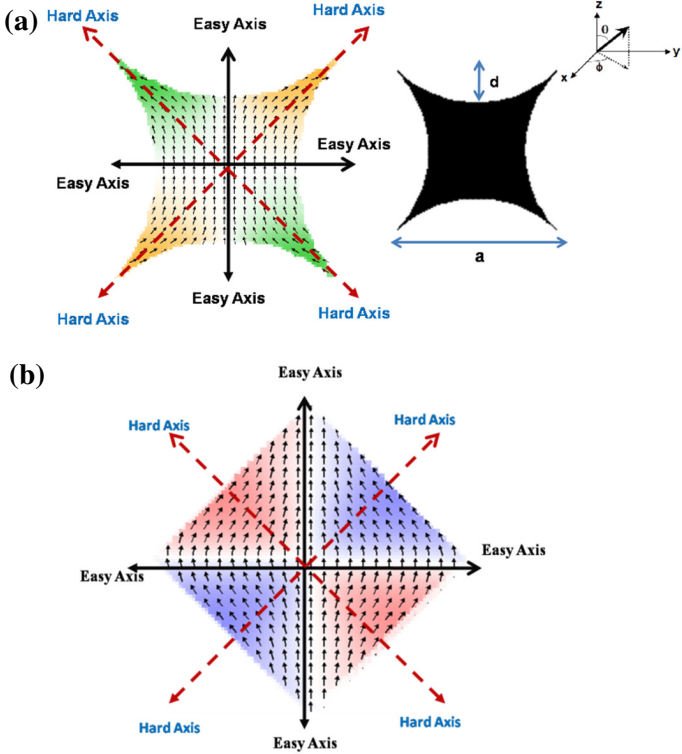

Fig. 1. Schematics of concave- and diamond-shaped four-state nanomagnets. The easy and hard axes of (a) concave-shaped nanomagnet with concavity, *d*, and lateral dimensions, *a*, and (b) diamond-shaped nanomagnet with internal diagonal length, *a*.

where $\Omega$ is the volume of the nanomagnet.

Considering stress, $\sigma = \varepsilon Y$, where $Y$ is the Young's modulus, $\varepsilon$ is the strain, and the stress is applied only along the y-axis, the stress anisotropy energy is:

$$U_{Stress-Anisotropy}(t) = -(\frac{3}{2}\lambda_S)\sigma(t)\Omega m_y^2(t) \quad (4)$$

where $\lambda_S$ is the magnetostrictive coefficient of the magnetic material. The effective field due to the stress anisotropy (Equation 4) is:

$$H_{eff-\sigma}(t) = (\frac{3}{\mu_o M_S}\lambda_S)\sigma(t)m_y(t) \quad (5)$$

This effective field is incorporated into the OOMMF simulations and has a direction along the y-axis since it was assumed that stress is only applied along this direction.

To analyze the magnetization reversal process and time evaluation of magnetic moment in the four-state diamond and concave nanomagnets, three dimensional (3D) micromagnetic simulations were performed using OOMMF. These OOMMF simulations perform time integration of the LLG equation, where the effective field includes the exchange, anisotropy, self-magnetostatic and external fields. The discretized cell size used in the simulations is 1 nm × 1 nm × 1 nm, implemented in the Cartesian coordinate system.

The parameters used for the Terfenol-D in the micromagnetic simulation are as follow: exchange constant, $A = 9 \times 10^{-12}$ J m$^{-1}$ [35], saturation magnetization, $M_s = 800$ kA m$^{-1}$, anisotropy constant, $K_1 = 0$ J m$^{-3}$ (no magnetocrystalline anisotropy), damping coefficient, $\alpha = 0.1$ [44], magnetostrictive coefficient $\frac{3}{2}\lambda_S = 900 \times 10^{-6}$, and Young's modulus, $Y = 80$ GPa [41]–[44].

## 3. Results

We investigate two different shapes: diamond and concave (shown in Fig. 1). In this work, Terfenol-D is selected as the magnetostrictive material for the four-state nanomagnets due to its high magnetostriction value, thereby requiring a lower amount of strain to switch its magnetization state. Here, we study the effects of both magnetic field and strain on the magnetization switching characteristics of these nanomagnets. In this section, the following characteristics are examined: (A) magnetization hysteresis (anisotropy field) (B) switching coherence, and (C) magnetization dynamics, in order to determine the best shape for coherent and reliable nanomagnet for future four-state memory and logic applications as well as for higher-order applications such as image recovery and recognition schemes [15]–[17].

### 3.1 Nonlinear Magnetization Hysteresis and Anisotropy Field

In order to determine the magnetization reversal process in the diamond and concave nanomagnets, micromagnetic simulations (OOMMF) were performed to verify its magnetization hysteresis. We study the hysteresis (m-B) loops of these nanomagnets for different thicknesses (10 nm and 15 nm) with lateral dimensions of 100 nm × 100 nm. The concavity depth, $d$, of the concave nanomagnet was chosen to be 20 nm. The results for both nanomagnets are shown in Fig. 2 which illustrates the normalized hysteresis loops for both shapes in the presence of an applied magnetic field along the $x$ (100) direction ($\phi = 0^o$).

The switching field for the diamond-shaped magnet with a thickness of 10 nm is ~16 mT. However, for a concave nanomagnet with the same lateral dimensions and thickness but having a concavity depth, $d = 20$ nm, this field increases to ~96 mT (Fig. 2a). When repeated for a thickness of 15 nm, we observe a switching field of 27 mT for the diamond nanomagnet and 141 mT for the concave nanomagnet. Therefore, the introduction of concavity to the sides of the diamond nanomagnet results in an increase in the switching field by a factor of ~6. This increase in the energy barrier between the easy and hard axes is associated with the coherent magnetization switching in the concave nanomagnets as opposed to the diamond nanomagnets. This phenomenon can be attributed to the configurational anisotropy introduced by the concavity in the sides of the nanomagnet.

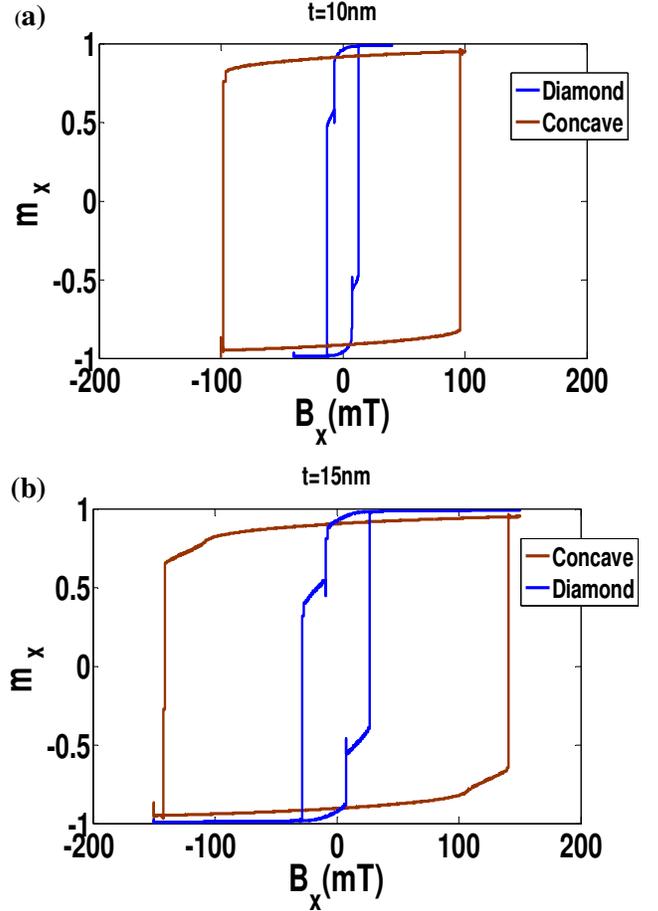

Fig. 2. Magnetization hysteresis (m-B) curves for the concave and diamond nanomagnets with dimensions of 100 nm × 100 nm and having a thickness of (a) 10 nm, and (b) 15 nm.

The anisotropy field for both diamond and concave nanomagnets was examined next, with the magnetization of each nanomagnet initialized in the +y direction, followed by the application of a magnetic field along the +x direction. Increasing the magnitude of the field in the +x direction causes the magnetization of the magnets to rotate, from the initial 'up' direction to the 'right' direction once the external magnetic field overcomes the energy barrier of the nanomagnet. The value of this field (that causes a 90° magnetization rotation) is taken as the anisotropy field of each nanomagnet. These simulations were performed for nanomagnets having the same lateral dimensions (100 nm × 100 nm) but different thickness and concavity depths, the results of which are illustrated in Fig. 3. For a diamond nanomagnet with a thickness of 8 nm, the anisotropy field is 6 mT. However, creating a concavity in its sides, with $d = 10$ nm, increases this anisotropy field to 15 mT, thereby increasing the energy barrier between the easy and hard axes by a factor of ~ 2.5. It is observed that the anisotropy field of the nanomagnets is sensitive to the thickness and concavity depth, with an increase in values of both parameters resulting in a corresponding increase in the anisotropy field.

It should be noted that increasing the thickness of the diamond magnet causes an increasing incoherence (vortex) in its switching characteristics (resulting in the double-jump hysteresis loop [31]), but not in the concave nanomagnets. The trend of low energy barrier values for the diamond nanomagnet persists till a thickness of 20 nm, above which the diamond nanomagnet shows an anisotropy field higher than that of a concave nanomagnet (for $d = 10$ nm) having the same lateral dimensions and thickness, as can be seen in Fig. 3. However, this increase in anisotropy field arises at the expense of increased vorticity in the magnetization.

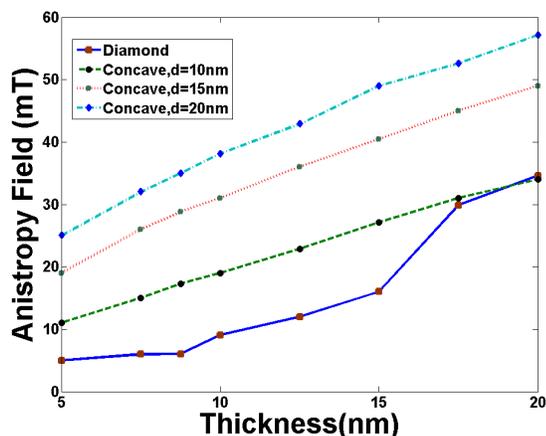

Fig. 3. Anisotropy field as a function of nanomagnet thickness for concave and diamond nanomagnets having lateral dimensions, $a = 100$ nm for different values of concavity.

*3.2 Magnetization vector patterns in Diamond- and Concave-Shaped Nanomagnets*

A single nanomagnet has two dominant and competing energy terms: (1) exchange energy, and (ii) anisotropy energy. In the previous section, it was shown that for higher thicknesses, the diamond nanomagnet shows a higher anisotropy field than that of a concave nanomagnet of similar dimensions (and concavity, $d = 10$ nm), at the expense of incoherent switching modes. Consequently, it is of interest to perform micromagnetic simulations using OOMMF and examine the evolution of this incoherence, from single-domain to incoherent vortex modes, in the diamond nanomagnet for different values of thickness as compared to that observed in a concave nanomagnet. Fig. 4 illustrates the magnetization patterns of a 100 nm × 100 nm diamond nanomagnet for various thicknesses.

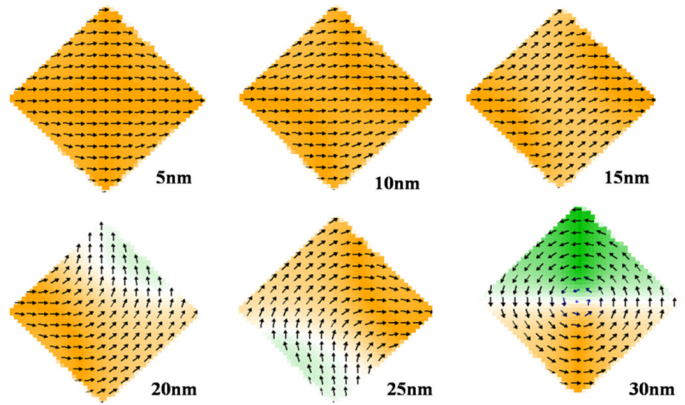

Fig. 4. Magnetization patterns of a 100 nm × 100 nm diamond nanomagnet for different thickness.

The magnetization of concave four-state nanomagnets with similar dimensions are also shown in Fig. 5 for concavity depths of 10 nm, 15 nm and 20 nm.

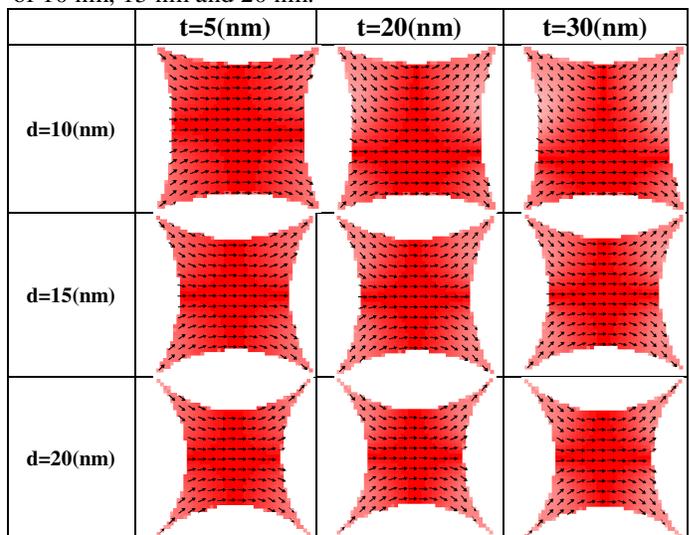

Fig.5. Magnetization patterns for concave nanomagnets of various thicknesses, t, and concavity depths, d.

It can be seen from Fig. 4 and Fig. 5 that while the diamond nanomagnets are susceptible to increased incoherent switching of the magnetization as the thickness increases, the concave nanomagnets show a trivial amount of incoherence (hence, the magnetization patterns of only three values of thickness – 5 nm, 20 nm, and 30 nm are shown). This insensitivity is prevalent even at larger thicknesses (with the same lateral dimensions). This phenomenon can be attributed to decreasing anisotropy energy in the concave nanomagnet which causes the exchange energy to dominate and results in coherent single domain magnetization. However, in sharp contrast, the diamond four-state nanomagnet demonstrates a higher degree of vortex states at higher thicknesses which can be attributed to increase in magnetostatic anisotropy energy. Figure 6 represents this phenomenon in terms of the incoherence percentage of the nanomagnets, calculated as the percentage of the magnetization vectors of single domain aligned along the +$x$ direction (final settled state) after a 90° rotation from the easy axis (+$y$ axis). Therefore, an incoherence percentage of 0% represents a complete rotation and settling of all magnetization moments vectors within a nanomagnet into easy axis along the +$x$ axis.

In the diamond nanomagnet, the incoherence percentage values are ~6.7% and 20% for a thickness of 5 nm and 10 nm (Fig. 6).

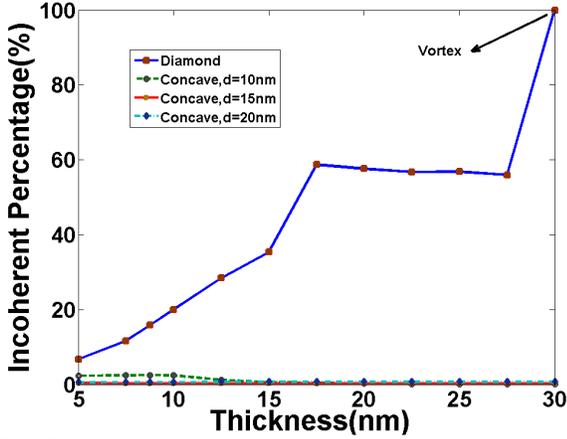

Fig. 6. The percentage of incoherent magnetic moments in diamond and concave nanomagnets vs. thickness.

However, for thicknesses greater than 17 nm, we can observe that this incoherence percentage value rises to ~60% resulting in a higher degree of magnetization vortex formation while for a thickness of 30 nm, the vortex state arises into 100% (no ferromagnetic ordering in the presence of no field).

*3.3 Magnetization Dynamics*

Thus far, micromagnetic simulation results studying the magnetization characteristics of diamond- and concave-shaped nanomagnets have shown that concave nanomagnets entail coherent magnetization switching modes with an incoherence percentage (vortex) rate that is near to zero, for a variety of thicknesses. The diamond nanomagnets, on the other hand, show increasing levels of vortex formation with increasing thickness. In this section, we investigate the time evolution of magnetization in these nanostructures using OOMMF in order to study the magnetization dynamics as the magnetization rotates from the hard axis and settles to its easy axis. The following two scenarios are examined. In Fig. 7(a), considering a (100 nm × 100 nm × 10 nm concave nanomagnet ($d$ = 20 nm), the initial magnetization was set along the hard axis ($\phi_0 = 45°$) with a 10° out-of-plane component, $\theta_0 = 80°$ (when $\theta = 90°$, the magnetization vector lies in the plane of the nanomagnet). The resulting torque generated, $|\vec{M} \times \vec{H}|$, causes the magnetization to rotate to the easy axis along the +$x$ direction, with a settling time of just ~0.5 ns. The magnetization dynamics is then examined for a diamond nanomagnet of similar dimensions, with its magnetization vector having the same initial configuration ($\phi_0 = 45°$, $\theta_0 = 80°$). The results, shown in Fig. 7(b), demonstrate an 'S' state switching mode (also with settling time of ~0.5 ns) resulting in an incoherence percentage of 20% in this diamond nanomagnet.

In order to demonstrate applications of these four-state nanomagnets for memory and logic, we have also explored the magnetization dynamics of (i) concave nanomagnets with a concavity of 10 nm, and (ii) four-state diamond nanomagnets (100 nm × 100 nm), both having a thickness of 6 nm, under the influence of mechanical stress (tension and compression). By applying a voltage across a thin-film piezoelectric layer grown on a substrate, an effective strain is generated that is transferred to magnetostrictive nanomagnets [39] fabricated on top of the thin film (Fig. 8a), resulting in magnetization rotation. This methodology of magnetization switching via strain is associated with low power memory and logic applications [9-11]. The schematic view of a multiferroic four-state nanomagnet and strain-based magnetization dynamics of the rotation are demonstrated in Figs. 8 and 9 for both tensile and compressive stresses of 20 MPa, with Terfenol-D possessing positive magnetostriction. In Fig. 8(a), the voltage across the pair of electrodes on the piezoelectric thin film layer implements both compressive and tensile strains. Fig. 8(b) demonstrates 90° coherent magnetization switching in a concave four-state nanomagnet from the initial 'up' direction to the 'right' under a *compressive* stress (negative electric field) of 20 MPa (Fig. 8b). Reverse switching (from 'right' to 'up') is also possible by applying a *tensile* stress (positive electric field) of 20 MPa along the same direction, as illustrated in Fig. 8(c). The magnetization is switched between easy axes with a stress of 20 MPa at a clock rate of 1.25 GHz, thus providing the capability for dense memory and logic applications. Terfenol-D has one of the highest magnetostriction coefficients, $(3/2)\lambda_S = 900 \times 10^{-6}$, among soft magnetic materials and Young's modulus of $Y$ = 80 GPa [41-44]. Thus, a relatively low amount of strain is needed for magnetization rotation in these single domain nanomagnets [10].

Similarly, Fig. 9(a) shows 90° magnetization switching from 'up' to 'right' in four-state diamond nanomagnet under a compressive stress of 20 MPa and reverse magnetization switching (from 'right' to 'up') is also performed by applying a substrate. Using a linear interpolation, the electric field required to produce a strain of $2.5 \times 10^{-4}$ is 0.5 MV/m. Therefore, the voltage to be applied to the top electrodes in Fig. 8a in order to generate a stress of 20 MPa is 0.5 MV/m × 100 nm = 50 mV. For the substrate thickness and electrode spacing, a uniform stress is generated in the region between the electrodes. We, therefore, assume uniform stress in the piezoelectric material for our OOMMF simulations. The energy dissipation associated with the application of this voltage (50 mV) between the top electrodes and bottom electrode, which acts as a capacitor [10], is $(1/2)\ CV^2$ for each electrode. If the dimensions of the top electrode pair are 120 nm × 120 nm, fabricated on top of the PZT thin film with a thickness of 100 nm, the associated capacitance, C = 1.275 fF, assuming the relative dielectric constant of the PZT thin film is 1000.

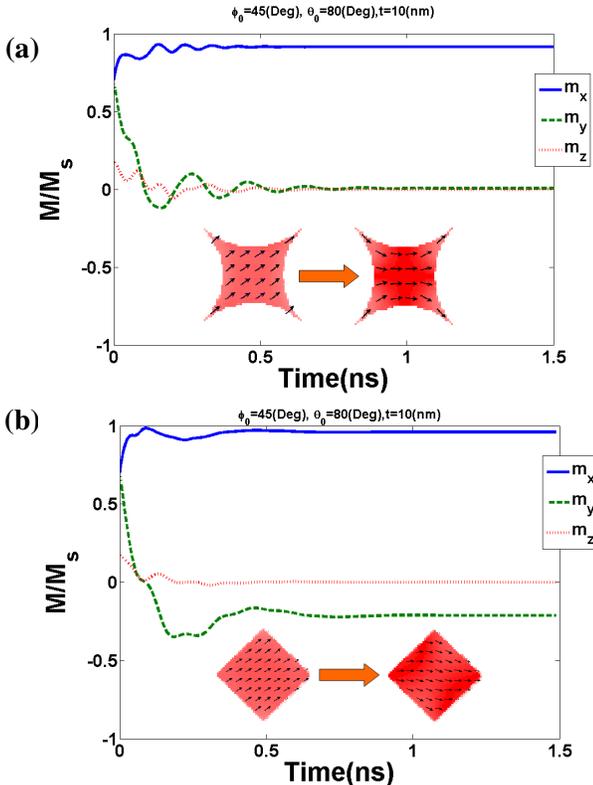

Fig. 7. Time evolution of the magnetization vector with an initial configuration of $\phi_0 = 45°$, $\theta_0 = 80°$ for (a) concave nanomagnet, and (b) diamond nanomagnet.

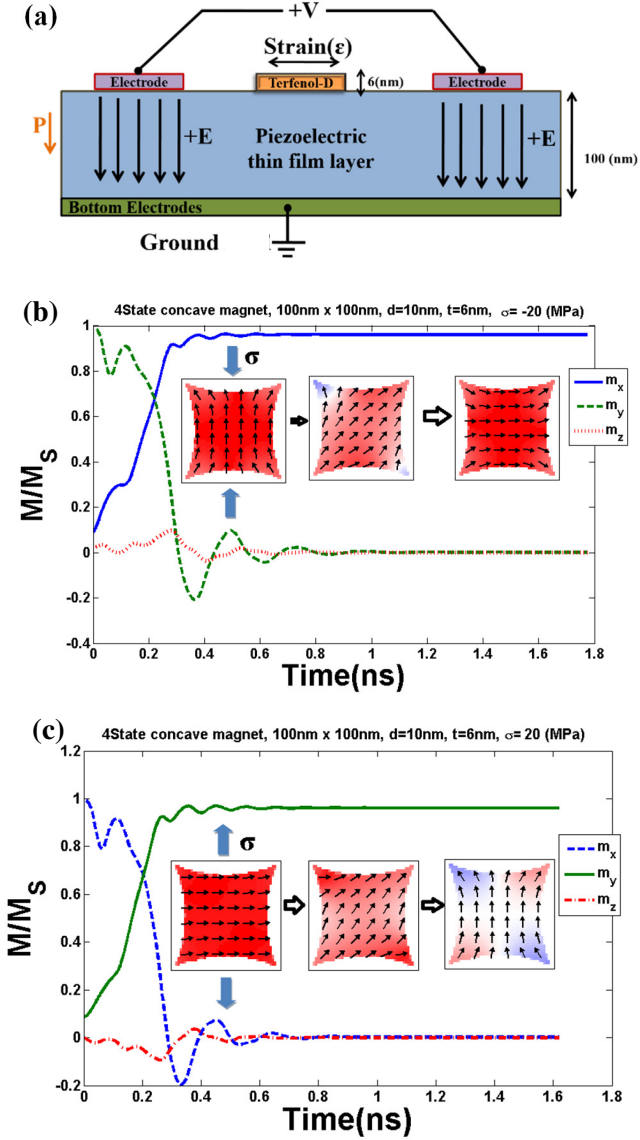

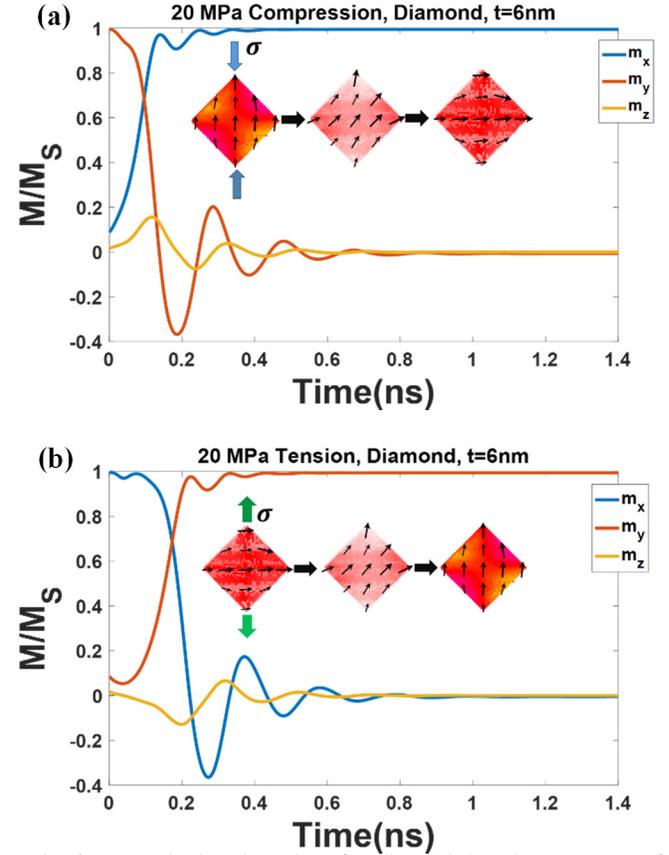

Fig. 8. (a) Cross-sectional schematic of a multiferroic four-state nanomagnet undergoing strain. Magnetization dynamics of a concave-shaped nanomagnet having dimensions of 100 nm × 100 nm, thickness = 6 nm, and concavity, d = 10 nm. (b) Magnetization switching from the initial 'up' to the 'right' state under a 20 MPa compressive stress at a clock rate of ~1.25 GHz, and (c) switching between 'right' to 'up' state with a 20 MPa tensile stress applied along the same axis (using the same electrodes).

Fig. 9. Magnetization dynamics of a diamond-shaped nanomagnet of dimensions 100 nm × 100 nm and thickness of 6 nm. (a) Magnetization switching from 'up' to 'right' because of 20 MPa compressive stress at a clock rate of 1.25 GHz, and (b) magnetization switching between 'right' to 'up' under 20 MPa tensile stress which generated with applying voltage to the same electrode for generating compressive stress (tensile stress is along the axis of the applied compressive stress).

tensile stress of 20 MPa along the same direction, as illustrated in Fig. 9(b). We note that in both cases, upon removal of stress at $t$ = 1 ns, there is negligible change in the magnetization state of the nanomagnets (compared to the settled magnetization state prior to stress removal).

In Figs. 8(b, c) and 9(a, b), we demonstrate the clock rate for writing information in both a concave four-state multiferroic nanomagnet with concavity depth, d = 10 nm, and a diamond nanomagnet (100 nm × 100 nm) with thickness of 6 nm is ~ 1.25 GHz under a stress of 20 MPa (tensile and compressive). This stress (20 MPa) produces a strain, $\varepsilon = 2.5 \times 10^{-4}$. Ref. [39] demonstrated that an electric field of 2 MV/m would produce a local strain of 1000 ppm through the use of a pair of electrodes across a micromagnetic disc on top of a PZT

Since the strain should be uniform across the four-state concave nanomagnet, the voltage (50 mV) should be applied simultaneously to both top electrodes. Therefore, the total energy dissipation ($E_d$) for both electrodes is $CV^2$. Consequently, applying a stress of 20 MPa on concave and diamond four-state nanomagnet results in $E_d$ = 3.187 aJ/bit, which is at least three orders of magnitude lesser than the energy dissipation associated with conventional spin-transfer-torque (STT-RAM) technology [40].

As concluded from the earlier section (Figs. 4–7) which studied magnetization switching in concave and diamond nanomagnets under the effect of a magnetic field (with no applied stress), the concave-shaped nanomagnets undergo coherent and more reliable magnetization switching even at a greater thickness, as compared to diamond-shaped nanomagnets which experience vortex formation during magnetization switching. Similarly, during stress-induced magnetization switching, the concave nanomagnets display a greater degree of robustness at higher thicknesses in terms of coherent magnetization switching, as compared to diamond-shaped nanomagnets. Figure 10 illustrates the magnetization switching in concave- and diamond-shaped nanomagnets under stress for a thickness of 16 nm (lateral dimensions for both structures are 100 nm × 100 nm; concavity depth in concave nanomagnet is 10 nm).

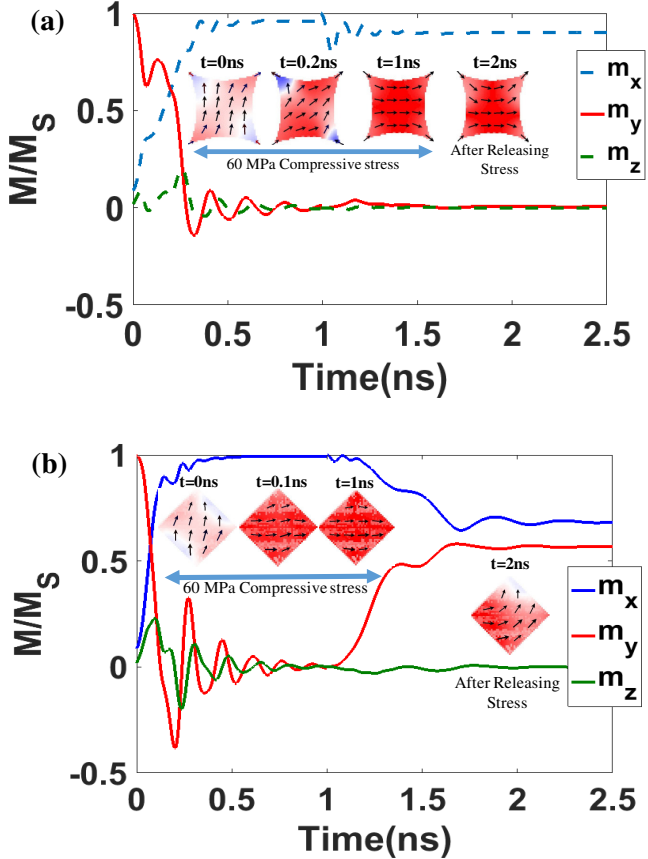

Fig. 10. Magnetization dynamics under a compressive stress of 60 MPa for (a) concave-shaped nanomagnet of dimensions 100 nm × 100 nm and thickness of 16 nm (concavity depth = 10 nm), showing switching from 'up' to 'right', and (b) diamond nanomagnet (100 nm × 100 nm, thickness = 16 nm) with magnetization switching from 'up' to 'right'. While the concave nanomagnet retains its magnetization state after removal of stress (t = 1 ns), the diamond nanomagnet experiences incoherency in switching and a higher degree of vorticity.

Due to the increased energy barrier resulting from the increase in thickness of the concave- and diamond-shaped nanomagnets, a greater stress of 60 MPa has to switch the magnetization. As illustrated in Fig. 10, this stress rotates the magnetization from the initial 'up' state to the 'right' in both nanomagnets. Upon removal of stress (at t = 1 ns), the concave nanomagnet retains its magnetization state (shown by the magnetization vectors at t = 2 ns) while the diamond nanomagnet develops an undesirable vortex-like state.

## 4. Discussion

Four-state nanomagnets possessing fourfold, symmetric anisotropy fields can be implemented in non-Boolean applications such as memory [36], [37], logic devices like four-state NOR gate [15] as well as in higher order applications such as image recognition and processing [17] and associative memory [38]. This study investigates the magnetization characteristics of a four-state diamond and concave nanomagnets and, in particular, the incoherent switching modes that arise as the thickness increases. Through shape engineering of the edges, concave nanomagnets are created and the subsequent deviation in the uniform magnetization would disappear. This effect is accompanied by coherent magnetization rotation (lower incoherence percentages as the concavity depth, $d$, increases), regardless of the nanomagnet thickness, thereby making concave-shaped nanomagnets more reliable than diamond nanomagnets during magnetization reversal (less susceptible to vortex state formation). Furthermore, it is important to emphasize the limitations associated with nanolithography when fabricating a precise diamond-shaped nanomagnet having sides of equal dimensions (100 nm). The preliminary experimental fabrication results show that a divergence of 15% from the nominal value results in the creation of a two-state, rather than the desired four-state, nanomagnet. Low-power strain-based magnetization reversal implementing these four-state nanomagnets in a multiferroic scheme is also explored for both concave and diamond nanomagnets with a low thickness of 6 nm. Both nanomagnets demonstrated similar magnetization dynamics at this thickness, however, increasing the thickness leads to a greater degree of vorticity in the diamond nanomagnets upon releasing the stress.

## 5. Conclusion

In this work, we have studied the influence of shape anisotropy on the single-domain magnetization states of four-state nanomagnets for two distinct shapes: (i) diamond, and (ii) concave, in the pursuit of reliable and efficient nanomagnets using magnetic field and strain-based switching. Various criteria were examined for these two shapes, such as size, magnetic hysteresis, concavity depth and thickness, in order to determine the ideal shape for coherent and reliable magnetization reversal for future magnetoelectronic devices. It was shown that concave nanomagnets acquire coherent and more reliable magnetization states (at higher thickness) while diamond nanomagnets are susceptible to incoherence due to increased vorticity in the magnetization states with increasing thickness. However, the concave nanomagnets of similar dimensions show little to no incoherence and are, in fact, quite robust to variations in thickness, a vital attribute in terms of fabrication of nanomagnetic devices. In addition, with the increasing interest in strain-based magnetization reversal for low-power, energy-efficient devices, we have demonstrated, for the first time, strain-clocked magnetization reversal in four-state concave and diamond nanomagnets with thickness of 6 nm (magnetization switching between easy axes states 'up' and 'right' and vice-versa) and possibility of writing information in this scheme with a clock rate of 1.25 GHz and an energy dissipation of less than 10 aJ. This study provides important insights into achieving reliable, coherent single-domain four-state nanomagnets, and a novel genre of a super energy-efficient magnetoelectric technology with the capability of possessing more information per nanomagnet. While nonvolatile devices offer both logic and memory capabilities, issues such as thermal fluctuations and inter-magnet coupling affect the resiliency, error probability and energy dissipation of these magnetic devices [45]–[47]. However, schemes using magneto-tunneling junctions (MTJs) such as the memory device proposed by Tiercelin et al. [48] and the straintronic-MTJ [49], implementing strain to switch the soft layer of the MTJ, have been studied for error-resilient information processing.

### Acknowledgment

This work is supported by the US National Science Foundation (NSF) under the SHF-Small grant CCF-1216614, NEB2020 grant ECCS-1124714 and by the Semiconductor Research Corporation (SRC) under NRI Task 2203.001.

The authors would like to acknowledge discussions with Prof. J. Atulasimha and Prof. S. Bandyopadhyay at the Virginia Commonwealth University on four-state nanomagnets and also Dr. M. Donahue at NIST for discussions on Object-Oriented Micromagnetic Framework (OOMMF) simulations.

### Appendix

In the main paper, we performed three-dimensional (3-D) micromagnetic simulation using OOMMF [29] to model the magnetization dynamics in a four-state concave multiferroic nanomagnet while subjected to strain. Since OOMMF cannot incorporate the effect of stress easily, the uniaxial anisotropy term was implied to induce magnetization switching by strain. The uniaxial anisotropy constant, $K_1 (J/m^3)$, was used instead of strain anisotropy, $K_\sigma$, and its direction was directed by uniaxial anisotropy along the [010] direction for implementing applied strain on the nanomagnet along the +$y$ direction, as illustrated in Fig. 8. The value of strain anisotropy constant is calculated as:

$$K_\sigma = \frac{3}{2}\lambda_S \sigma \ (J/m^3) \quad (A1)$$

To verify the accuracy of the micromagnetic OOMMF simulations for strain induced magnetization switching in a four-state concave nanomagnet, we compared OOMMF simulations for an ellipsoid nanomagnet subjected to stress with the macrospin approximation (Landau-Lifshitz-Gilbert formalism or LLG) in equation 2. To perform this benchmarking, a single domain ellipsoid nanomagnet with dimensions of 100nm × 90nm × 10nm was used. Terfenol-D was considered as the magnetic material of this ellipsoid nanomagnet with the same properties as in the main paper.

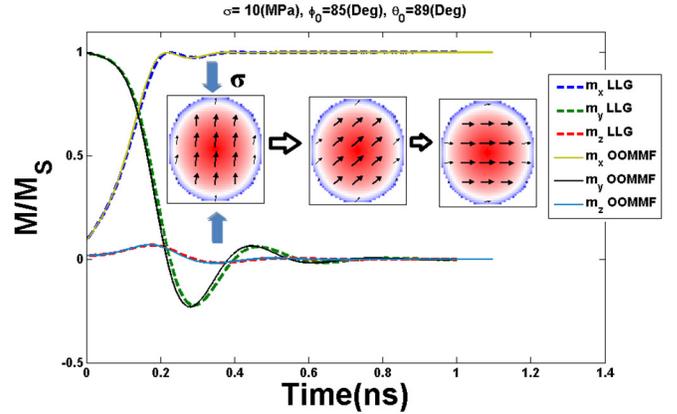

Fig. A1: Comparison of the OOMMF results with LLG simulation based on applying 10MPa strain on the ellipsoid nanomagnet.

The applied compressive stress on the ellipsoid nanomagnet was -10 MPa which gives a uniaxial anisotropy, $K_\sigma = -9000 \ J/m^3$. The initial magnetization direction was $\varphi_0 = 85°$ and $\theta_0 = 89°$. After applying a compressive stress of 10 MPa on the ellipsoid nanomagnet, its magnetization underwent a rotation from the easy axes (up) to the hard axis (right) within ~0.6 ns (Fig. A1). The micromagnetic calculation is initialized with uniform magnetization across the ellipsoid nanomagnet and then relaxed to its equilibrium configuration. The simulation was also terminated when the normalized residual torque satisfied convergence of the $|\vec{m} \times \vec{H}|$ to a value less than $10^{-5}$. Figure A1 shows both LLG and OOMMF simulations for an ellipsoid nanomagnet which was subjected to a compressive stress of 10 MPa, showing a high degree of agreement. The overall magnetization switching behavior obtained from both LLG and OOMMF are very similar which validate the accuracy of micromagnetic simulations on strain-clocked magnetization of concave four-state nanomagnets in this work.

### References

[1] S. A. Wolf, D. D. Awschalom, R. a Buhrman, J. M. Daughton, S. von Molnár, M. L. Roukes, a Y. Chtchelkanova, and D. M. Treger, "Spintronics: a spin-based electronics vision for the future.," *Science*, vol. 294, no. 5546, pp. 1488–95, Nov. 2001.

[2] G. Csaba, A. Imre, G. H. Bernstein, W. Porod, and V. Metlushko, "Nanocomputing by field-coupled nanomagnets," *IEEE Trans. Nanotechnol.*, vol. 1, no. 4, pp. 209–213, Dec. 2002.

[3] R. P. Cowburn and M. E. Welland, "Room Temperature Magnetic Quantum Cellular Automata," *Science (80-. ).*, vol. 287, no. 5457, pp. 1466–1468, Feb. 2000.

[4] D. Carlton, N. Emley, E. Tuchfeld, and J. Bokor, "Simulation Studies of Nanomagnet-Based Logic Architecture," *Nano Lett.*, vol. 8, no. 12, pp. 4173–4178, 2008.

[5] S. Bandyopadhyay and M. Cahay, "Electron spin for classical information processing: a brief survey of spin-based logic devices, gates and circuits.," *Nanotechnology*, vol. 20, no. 41, p. 412001, Oct. 2009.

[6] S. Salahuddin and S. Datta, "Interacting systems for self-correcting low power switching," *Appl. Phys. Lett.*, vol. 90, no. 9, p. 093503, 2007.

[7] R. P. Cowburn, D. K. Koltsov, A. O. Adeyeye, M. E. Welland, and D. M. Tricker, "Single-Domain Circular Nanomagnets," *Phys.Rev.Lett.*, vol. 83, no. 5, pp. 1042–1045, Aug. 1999.

[8] A. Imre, G. Csaba, L. Ji, A. Orlov, G. H. Bernstein, and W. Porod, "Majority logic gate for magnetic quantum-dot cellular automata.," *Science*, vol. 311, no. 5758, pp. 205–8, Jan. 2006.

[9] J. Atulasimha and S. Bandyopadhyay, "Bennett clocking of nanomagnetic logic using multiferroic single-domain nanomagnets," *Appl. Phys. Lett.*, vol. 97, p. 173105, 2010.

[10] M. Salehi Fashami, K. Roy, J. Atulasimha, and S. Bandyopadhyay, "Magnetization dynamics, Bennett clocking and associated energy dissipation in multiferroic logic.," *Nanotechnology*, vol. 22, no. 15, p. 155201, Apr. 2011.

[11] M. Salehi Fashami, J. Atulasimha, and S. Bandyopadhyay, "Magnetization dynamics, throughput and energy dissipation in a universal multiferroic nanomagnetic logic gate with fan-in and fan-out.," *Nanotechnology*, vol. 23, no. 10, p. 105201, Mar. 2012.

[12] F. J. Albert, J. A. Katine, R. A. Buhrman, and D. C. Ralph, "Spin-polarized current switching of a Co thin film nanomagnet," *Appl. Phys. Lett.*, vol. 77, no. 23, p. 3809, 2000.

[13] K. Roy, S. Bandyopadhyay, and J. Atulasimha, "Hybrid spintronics and straintronics: A magnetic technology for ultra low energy computing and signal processing," *Appl. Phys. Lett.*, vol. 99, no. 6, p. 063108, 2011.

[14] K. Roy, S. Bandyopadhyay, and J. Atulasimha, "Energy dissipation and switching delay in stress-induced switching of multiferroic nanomagnets in the presence of thermal fluctuations," *J. Appl. Phys.*, vol. 112, no. 2, p. 023914, Jul. 2012.

[15] N. D'Souza, J. Atulasimha, and S. Bandyopadhyay, "Four-state nanomagnetic logic using multiferroics," *J. Phys. D. Appl. Phys.*, vol. 44, no. 26, p. 265001, Jul. 2011.

[16] N. D'Souza, J. Atulasimha, and S. Bandyopadhyay, "An energy-efficient Bennett clocking scheme for 4-state multiferroic logic," *Nanotechnology, IEEE Trans.*, vol. 11, no. 2, pp. 418–425, 2012.

[17] N. D'Souza, J. Atulasimha, and S. Bandyopadhyay, "An ultrafast image recovery and recognition system implemented with nanomagnets possessing biaxial magnetocrystalline anisotropy," *Nanotechnology, IEEE Trans.*, vol. 11, no. 5, pp. 896–901, 2012.

[18] A. Imre, G. Csaba, G. H. Bernstein, W. Porod, and V. Metlushko, "Investigation of shape-dependent switching of coupled nanomagnets," *Superlattices Microstruct.*, vol. 34, no. 3–6, pp. 513–518, Dec. 2003.

[19] R. Naik, C. Kota, J. Payson, and G. Dunifer, "Ferromagnetic-resonance studies of epitaxial Ni, Co, and Fe films grown on Cu(100)/Si(100).," *Phys. Rev. B. Condens. Matter*, vol. 48, no. 2, pp. 1008–1013, Jul. 1993.

[20] P. P. Chow, "Molecular Beam Epitaxy," in *Thin film processes II*, J. L. Vossen and W. Kern, Eds. Boston: Academic Press, 1991.

[21] E. Boyd, "Magnetic anisotropy in single-crystal thin films," *IBM J. Res. Dev.*, no. April, pp. 116–129, 1960.

[22] C. P. Wang, "A Coupled Magnetic Film Device for Associative Memories," *J. Appl. Phys.*, vol. 39, no. 2, p. 1220, 1968.

[23] W. T. Siegle, "Exchange Coupling of Uniaxial Magnetic Thin Films," *J. Appl. Phys.*, vol. 36, no. 3, p. 1116, 1965.

[24] F. Lee, "Shape-induced biaxial anisotropy in thin magnetic films," *IEEE Trans. Magn.*, vol. 4, no. 3, pp. 502–506, Sep. 1968.

[25] R. P. Cowburn, D. K. Koltsov, A. O. Adeyeye, and M. E. Welland, "Designing nanostructured magnetic materials by symmetry," *Europhys. Lett.*, vol. 48, no. 2, pp. 221–227, Oct. 1999.

[26] P. Vavassori, D. Bisero, F. Carace, A. di Bona, G. Gazzadi, M. Liberati, and S. Valeri, "Interplay between magnetocrystalline and configurational anisotropies in Fe(001) square nanostructures," *Phys. Rev. B*, vol. 72, no. 5, p. 054405, Aug. 2005.

[27] B. Lambson, Z. Gu, D. Carlton, S. Dhuey, A. Scholl, A. Doran, A. Young, and J. Bokor, "Cascade-like signal propagation in chains of concave nanomagnets," *Appl. Phys. Lett.*, vol. 100, no. 15, p. 152406, 2012.

[28] B. Lambson, Z. Gu, M. Monroe, S. Dhuey, A. Scholl, and J. Bokor, "Concave nanomagnets: investigation of anisotropy properties and applications to nanomagnetic logic," *Appl. Phys. A*, vol. 111, no. 2, pp. 413–421, Mar. 2013.

[29] M. J. Donahue and D. G. Porter, "OOMMF User's Guide, Version 1.0, Interagency Report NISTIR 6376," 1999.

[30] R. P. Cowburn, A. Adeyeye, and M. Welland, "Configurational Anisotropy in Nanomagnets," *Phys. Rev. Lett.*, vol. 81, no. 24, pp. 5414–5417, Dec. 1998.

[31] L. Torres, E. Martinez, L. Lopez-Diaz, and J. Iñiguez, "Micromagnetic switching of patterned square magnetic nanostructures," *J. Appl. Phys.*, vol. 89, no. 11, p. 7585, Jun. 2001.

[32] R. P. Cowburn and M. Welland, "Micromagnetics of the single-domain state of square ferromagnetic nanostructures," *Phys. Rev. B*, no. October, pp. 9217–9226, 1998.

[33] I. D. Mayergoyz, G. Bertotti, and C. Serpico, *Nonlinear Magnetization Dynamics in Nanosystems*, 1st ed. Elsevier Science, 2009, p. 480.

[34] T. L. Gilbert, "Classics in Magnetics A Phenomenological Theory of Damping in Ferromagnetic Materials," *IEEE Trans. Magn.*, vol. 40, no. 6, pp. 3443–3449, Nov. 2004.

[35] G. Dewar, "Effect of the large magnetostriction of Terfenol-D on microwave transmission," *J. Appl. Phys.*, vol. 81, no. 8, p. 5713, 1997.

[36] Z. Shi, C. Wang, X. Liu, and C. Nan, "A four-state memory cell based on magnetoelectric composite," *Chinese Sci. Bull.*, vol. 53, no. 14, pp. 2135–2138, Jul. 2008.

[37] T. Uemura, T. Marukame, K. Matsuda, and M. Yamamoto, "Four-State Magnetoresistance in Epitaxial CoFe-Based Magnetic Tunnel Junctions," *IEEE Trans. Magn.*, vol. 43, no. 6, pp. 2791–2793, Jun. 2007.

[38] V. Roychowdhury, D. Janes, S. Bandyopadhyay, and X. Wang, "Collective computational activity in self-assembled arrays of quantum dots: a novel neuromorphic architecture for nanoelectronics," *Electron Devices, IEEE Trans.*, vol. 43, no. 10, pp. 1688–1699, 2002.

[39] J .Cui, J. L. Hockel, P. K. Nordeen, D. M. Pisani, C. Liang, G. P. Carman, and C. S. Lynch. "A method to control magnetism in individual strain-

mediated magnetoelectric islands." Applied Physics Letters 103, no. 23 (2013): 232905.

[40] K. L.Wang, J. G. Alzate, and P. Khalili-Amiri. "Low-power non-volatile spintronic memory: STT-RAM and beyond." Journal of Physics D: Applied Physics 46, no. 7 (2013): 074003.

[41] Chikazumi S 1964 Physics of Magnetism (New York: Wiley)

[42] Ried K, Schnell M, Schatz F, Hirscher M, Ludescher B, Sigle W and Kronmuller H 1998 Phys. Status Solidi a 167 195 Abbundi R and Clark A E 1977 IEEE Trans. Mag. 13 1547

[43] Kellogg R A and Flatau A B 2008 J. Intell. Mater. Syst. Struct. 19 583

[44] Walowski J, Djorjevic Kaufman M, Lenk B, Hamann C, McCord J and M¨unzenbergM 2008 J. Phys. D: Appl. Phys. 41 164016

[45] M. Salehi-Fashami, K. Munira, S. Bandyopadhyay, A. W. Ghosh and J. Atulasimha. "Switching of dipole coupled multiferroic nanomagnets in the presence of thermal noise: reliability analysis of hybrid spintronic-straintronic nanomagnetic logic" *IEEE Trans. Nanotech,* vol. 12, pp. 1206–1212, 2013.

[46] F. M. Spedalieri, A. P. Jacob, D. E. Nikonov & V. P. Roychowdhury. "Performance of magnetic quantum cellular automata and limitations due to thermal noise." *IEEE Trans. Nanotech.* vol. 10, pp. 537–546, 2011.

[47] F. A. Shah, G. Csaba, M.T. Niemier, X.S. Hu, W. Porod, G.H. Bernstein, Error analysis for ultra dense nanomagnet logic circuits, J. Appl. Phys. 117 (2015) 17A906.

[48] N. Tiercelin, Y. Dusch, A. Klimov, S. Giordano, V. Preobrazhensky, P. Pernod, Room temperature magnetoelectric memory cell using stress-mediated magnetoelastic switching in nanostructured multilayers, Appl. Phys. Lett. 99 (2011) 192507.

[49] A. K. Biswas, J. Atulasimha, and S. Bandyopadhyay. " An error-resilient non-volatile magneto-elastic universal logic gate with ultralow energy-delay product." *Scientific Reports*, vol. 4, pp. 7553, 2014.